# A Theoretical Approach to initiate Mobile Assisted Language Learning among school leavers and University Students of Sri Lanka

**Abstract**

Mobile devices have proven to be an appropriate tool in the area of mobile language learning, by expanding the horizons of learning beyond the classroom education. Power of mobile devices and advanced technologies have made ease of access to educational resources and materials to the learner. This study is a part of an ongoing research focusing on enhancing English Language Learning among school leavers and undergraduates. The objective of this paper is to present the results of a study conducted to identify the prospects of m-learning among the target study community, who were school leavers and undergraduates of Sri Lanka. A design based research methodology will be used in this study along with activity theory as a framework in order to analyse the situation and find out the prospects in introducing m-learning among the target community.

## 1 Introduction

Mobile learning has been found to be one of the major developing areas in recent years in educational field. The advancement of mobile technologies has also paved way for new learning experience through mobile based learning opportunities. The growing number of researches in the area of mobile language learning indicates the acceptance of m-learning among the community (Aamri and Sulaiman 2011; Corbail and Corbeil 2007; Gounder 2011; Kim, Mims and Holmes 2006).

In Sri Lanka the low proficiency of English Language has been a rising issue among the young adults. Many programs have been carried out to enhance the English proficiency which have been minimally successful comparing to the investments made on it (Karunaratne 2003). One of the evidence that depicts the issue with English Proficiency is the very low and declining English Language Proficiency Index which is rated by 'Education First, English Language Proficiency Index (EF-EPI). Sri Lanka has a declining index throughout last few years (EF - EPI 2015). Further, studies reveal that many qualified youth in Sri Lanka tend to possess the 'technical skills' or academic qualification for a job but lack the soft skills to convert knowledge into a profession (Ratwatte 2011). Following this problem, the increased mobile penetration and the capabilities of mobile devices (Gounder 2011) are evident as opportunities by the notion of mLearning.

This study is a part of an ongoing research which focuses on identifying the effectiveness of mobile learning among school leavers in the area of English Language Leaning and this paper has been written based on the initial stage of the research which is requirement elicitation and analysis phase. Novelty of this work lies as such kind of study has not been conducted among the given target group in the context of Sri Lanka, since in Sri Lanka m-learning is still in the phase of its inception. The results of this study can be used to design and develop a mobile learning application suiting these learner communities, which will ensure the sustainability of the application. The results of this study proves that there is a strong need for English Language Learning among the respondents and they are having a strong positive attitude towards m-learning. Hence it depicts that it is worth to continue this research by providing suitable means of mobile learning lessons among the target community.

Section 1 provides an introduction to the study, the 2nd Section includes a brief literature review of the research which have been carried out in the area of Mobile Learning, Activity theory and mobile learning in Sri Lanka. The 3rd section presents the methodology used to conduct this study and continued by findings and analysis in the 4th section. The 5th section presents the need of mobile learning in the Sri Lankan context, based on the results found in the previous section.

## 2 Literature Review

### 2.1 Mobile Learning

Many authors have perceived m-learning in different way, and they have come up with various definitions for the term mobile learning. Traxler has stated that there are number of studies in this emerging field, but the concept of mobile learning is still unclear (Yang 2013). One of the obvious and simplest definition of mLearning is, "learning delivered through the use of mobile technologies or devices". This definition mainly focuses on technological aspect of mobile learning rather than the pedagogical aspect (Traxler 2007). Osman (2010) has defined mobile learning as "any type of learning that takes place in learning environments and spaces that take account of the mobility of technology, mobility of the learner and mobility of learning". But as a whole, Laouriss and Eteokleous have conducted a research in drawing an educationally relevant definition for mobile learning (Laouris & Eteokleous). They have analysed the existing definitions by various authors and have derived that mLearning can be defined in the context of devices, in the context of the learning environment and learning experiences and as a function of its facets.

## 2.2 Activity Theory

Activity theory is a theoretical framework for analysing human practices as developmental processes with both individual and social levels interlinked at the same time (Kutti 1995). Activity theory has been found as a suitable tool to analyse the m-learning contexts as well (Uden 2007). Below is the basic structure of an activity system.

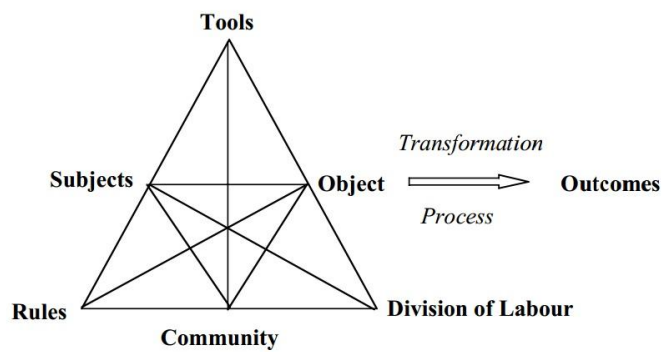

*Figure 1: The Activity Triangle Model (Engestrom 1987)*

An activity system comprises of Subjects, Object and Community components along with the mediators of Tools, Rules and Division of Labour. The 'Object' component reflects the motivational or purposeful nature of human activity that allows humans to control their own behaviour (Mwanza 2001). A subject can be an individual or group who are engaged in an activity. An activity is that, carried out by a subject in order to achieve an object (Objective) which can be mediated by tools. An object will be transformed into an outcome. The other components of the framework comprises of the rules, community and rules of labour, which denotes are related to a collective activity (Uden 2007). Community is made up of one or more people sharing the same objective. The rules mediates the relationship between the subjects and community. The division of labour denotes the way how tasks are divided between the community members, thus it mediates the relationship between the community and object (Uden 2007).

## 2.3 Mobile based Language learning in Sri Lanka

According to the Telecommunication Regulation Commission of Sri Lanka, the mobile connections are increasing dramatically and have reached 21,394,262 subscribers in March 2014. Meanwhile the Mobile Broadband connections also have increased to 1,777,955, which is a 6% more than year 2013 (Telecommunications Regulatory Commission of Sri Lanka 2015).

With the advancements in the technology and the penetration of the high end mobile devices, it is no wonder that Sri Lanka also has certain mobile learning initiatives. Dialog (Dialog Sri Lanka 2015) and Mobitel (Mobitel Sri Lanka 2015) which are two major mobile telecommunication service providers of Sri Lanka are already providing mLearning solutions through their products. Further, British council (British Council Sri Lanka 2015) also have their own SMS based lessons and mobile applications to enhance English Language Learning. The existing m-learning initiatives were provided by business organizations and had their own business line, which also incurred a cost for the user at a point.

# 3 Methodology

The entire research adopts the Design Based Research (DBR) methodology since it has been found as one of the appropriate research methods in exploring possibilities for creating novel learning and teaching environments (DBRC 2003). DBR is an emerging paradigm for the study of learning in context through the systematic design and study of instructional strategies and tools. DBR focus in solving current real-world problems by designing and enacting interventions as well as extending theories and refining design principles (DBRC 2003). It is grounded in both theory and the real-world context and in terms of research process it is interactive, iterative and flexible (Wang and Hannafin 2005). The phases of identified in the whole study were requirement gathering and elicitation, design of the learning solution, development and testing of the application and evaluation.

This paper was written based on the initial stage of the study which was based on identifying the prospects of m-learning among the target learners through the requirement gathering and elicitation. We use Activity theory as a framework in order to analyse the background and identify the prospects in introducing m-learning among the target community.

An online questionnaire survey was conducted among the target community with open and closed end questions in order to collect quantitative and qualitative data through adopting a mixed method research approach. The objectives of this survey were 1) To collect and analyse the data related to the demographics of learners, their interaction patterns and their attitude towards m-learning 2) Identify the mobile penetration and the mobile usage patterns among the target learners 3) Identify the problems related to English Language Learning among the target learners.

The user groups are selected from School leavers and university students. The target community have been divided into two different groups based on their interaction and exposure to virtual learning environments. The two learner groups are identified as Students with strong face to face interaction (G1) and Students who are having weak face to face interaction (G2). G1 comprised of internal undergraduates, since it is assumed that the face to face interaction will be higher among them. Further, the samples of G2 were selected from the external undergraduates since they mostly interact through the virtual environment.

The survey consisting of mixed methods of questions to gain statistical responses. 214 responses were collected from both groups within a period of 4 weeks. Out of the whole set of students who participated in the survey, 117 were from internal undergraduate programme and rest of 97 were from the external undergraduates.

The data was analysed using the Statistical Package for Social Sciences (SPSS) software version 20.0. Activity theory was used as a framework to analyse situation and find out the prospects of m-learning among the study group.

## 4   Findings and Analysis

This section will discuss the questionnaire results received from the respondents. The prospects of introducing m-learning will be discussed through this section by using Activity theory as a theoretical framework.

### 4.1   Activity Theory Based Analysis

The responses which were received was analysed in terms of the subjects, tools, objects, community, rules and division of labour which are the typical components of the activity system. The following figure depicts the application of the activity system.

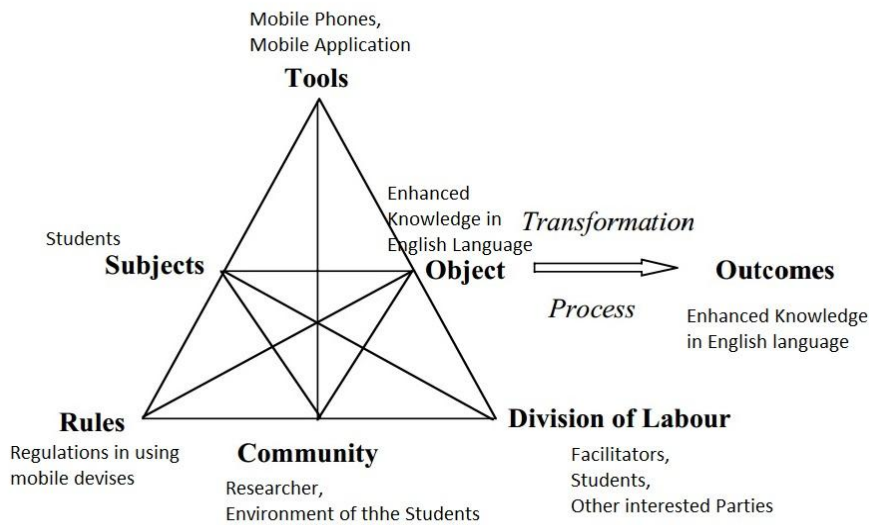

*Figure 2: Activity system for the m-learning application*

As the survey mainly focused on collecting data with respect to learner demographics, interaction patterns, learner attitude towards m-learning, mobile usage patterns and possible mobile solutions and the problems related to English Language Learning. Hence the activity system will be used only to discuss the learner related data and mobile usage related data through the activity system with respect to subjects and tools.

### 4.1.1 Subject Analysis

The final sample comprised of 214 responses from both learner groups who were internal undergraduates (G1) and External Undergraduates (G2). The analysis revealed that the learners had a strong need to enhance their English proficiency and positive attitude towards m-learning. Following table depicts the characteristics of the learner.

| Characteristics | | Number of respondents = 214 | |
|---|---|---|---|
| | | Number | Percentage |
| Age | 18 – 25 | 183 | 85.5% |
| | Above 25 | 31 | 14.5% |
| Gender | Male | 146 | 68% |
| | Female | 68 | 32% |
| Owning a mobile phone | Yes | 214 | 100% |
| | No | 0 | - |
| Type of mobile phone | Smart phone/ Tab | 153 | 71% |
| | A phone with basic features | 61 | 29% |
| Prospects of purchasing a smart phone (Issued among those who does not own a smart phone (61)) | Yes, in the near future | 30 | 49% |
| | Yes, don't know when | 23 | 38% |
| | Not interested in | 8 | 13% |

| | using a smartphone | | |
|---|---|---|---|

*Table 1. Characteristics of the learners*

Out of the whole respondents, 68% were Male and rest of 32% were female. Since most of the respondents were undergraduates, majority of them were within the age group of 18 – 25. There were 31 respondents who were above 25 years old, who were found in the group of external undergraduate program. Further, it was revealed that each and every student own a mobile thus the mobile ownership was 100%. The question issued with the objective of finding the usage of smartphone applications revealed that 71% of the respondents own a smartphone and the prospects of moving to a smart phone is higher. The question issued among the non-smartphone users about purchasing a smartphone, revealed that 87% of the non-smartphone users are planning to purchase a smart phone, which could be recognised as a significant number.

### 4.1.2 Analysis of the tools

Tools mediates the subject achieving the objective. In this study, the tools were identified as mobile device and the technologies. Following table shows the analysis of the tools and technologies among the respondents. The values are given in percentages for each learner groups.

| Groups | Characteristics | | Percentage (%) | |
|---|---|---|---|---|
| | | | G1 | G2 |
| Smartphone users | Operating system | Android | 47.1 | 37.4 |
| | | iOS | 2.6 | 4.5 |
| | | Symbian | 1.3 | 0.6 |
| | | Windows Mobile | 3.2 | 1.9 |
| | | Blackberry | 0 | 1.3 |
| | Familiarity of installing mobile apps | Yes | 53.6 | 39.9 |
| | | No | 1.3 | 5.2 |
| | Availability of English Language Learning applications | Yes, frequently used | 13.7 | 9.1 |
| | | Yes, Rarely used | 16.7 | 14.4 |
| | | No | 24.2 | 21.6 |
| All respondents | Use of phones for internet browsing | Yes, frequently | 43 | 27.1 |
| | | Yes, rarely | 8.4 | 10.7 |
| | | Not interested | 0 | 0.9 |
| | | Feature not available | 2.8 | 4.7 |
| | | Not aware of the feature | 0.5 | 1.9 |
| | Frequently Accessed Sites | Social networks | 44.9 | 30 |
| | | Online dictionaries and encyclopaedias | 18.2 | 17.3 |
| | | Web search engines | 43 | 33.2 |
| | | Other | 1.9 | 3.2 |

| | | | |
|---|---|---|---|
| Preferred method of using mobile for English Language Learning | Browse learning content (Do activities) | 32.2 | 33.6 |
| | Take small Tests | 16.8 | 26.6 |
| | Listen to audios | 22.9 | 34.1 |
| | I do not like to use the mobile to learn English | 3.2 | 1.9 |
| Intention to use mobile to learn English | Yes | 29.4 | 22.9 |
| | No | 25.2 | 22.4 |

*Table 2. Analysis of the tools*

Almost all of the respondents who owned a smartphone had android as the major operating system. It was identified that respondents are mostly using the features such as SMS, social networking, email, radio and camera. It is also noted that the other features such as calendar, maps, games, dictionaries and reading documents are also used by considerable amount of respondents. Further, through this results it is evident that the group of non-smart phone users are also accessing internet through their java enable phones. In the group of non-smart users, SMS was found to be more popular. Further, there was a considerable amount of users for radio, internet, camera, educational features and games as well.

Further, the findings show that the intention to use mobile to learn English is marginal with a slight increase of positive responses. Higher number of respondents are interested to receive test results through mobile (80%), which shows a promising attitude towards m-learning.

## 4.2 The attitude towards m-learning

When analysing the qualitative data it was evident that there is a strong need to enhance English proficiency among the adult learners. Respondents had issues with English language learning and some of the common issues were identified in the areas of Vocabulary, Speaking, Grammar, Writing and Listening. Some of the main reasons for the issues were, fear to practice in public, lack of proper resources, influence of the native language in day to day activities, lack of guidance, personal issues such as time management and motivational issues.

Respondents were provided with 5 point likert scale where each choice were ranked as 'Strongly Agree', 'Agree', 'Neutral', 'Disagree' and 'Strongly Disagree'. The 'Strongly Agree' and 'Agree' responses of the Positive statements and the 'Strongly Disagree' and 'Disagree' responses of the negative statements were analysed and the correlations of those statements between the two groups were identified.

| | Percentage | |
|---|---|---|
| | G1 | G2 |
| Use of mobile device to improve English Knowledge | 48.6% | 44.6% |
| Mobile learning as a flexible mode of learning | 48.6 % | 41.1 % |
| Ready to invest in a better device | 23.8% | 21.0% |
| Spending for mobile internet connection | 38.8 % | 29.4% |
| Perception towards improving English speaking skills through mobile devices | 40.2% | 38.3% |
| Ready to install mobile apps that improve English knowledge | 45.8% | 42.1% |
| Keep learning materials in the mobile | 50.9% | 42.1% |
| device | | |
| M-Learning helps to utilize the time productively | 45.3% | 41.1% |
| Enhanced teacher student communication through m-Learning | 28.5% | 37.8% |
| M-Learning will be advantageous | 41.6% | 32.7% |
| Possibility of Having team work through mobile | 20.6% | 15.6 % |
| Cost incurred through mobile devices | 18.2% | 15% |

*Table 3. Attitude towards m-learning*

As a whole, this question revealed that the respondents have strong positive attitude towards m-learning. The first statement was about the learner perception on using mobile devices for learning English and through this it is evident that more than 90% of the respondent believe that mobile devices can be used to enhance English Language Learning through mobile. M-learning is highly acceptable due to its flexibility and portability. It is confirmed through the second statement where around 90% of the respondents were advocates of m-learning due to its flexibility. Further, cost is one of the restraining factors in the case of m-learning and this has been once again proved through this survey. Most of the responses were found to be neutral when inquired about the cost factors. 31 % provided neutral response where 45% was towards positive response. Rest of 24% was towards negative response. This depicts that the cost is a major factor to be considered. When inquiring about the cost associated to the internet access, majority of the respondents showed a neutral to positive attitude. Around 68% of the respondents are ready to pay for the internet connection where 8% are more concerned towards the data connection cost. Rest of 24% have provided a neutral response. Further, 97% of the respondent are having neutral to positive attitude in using the mobile phones to enhance speaking skills. A statement was issued in order to test whether the learners are willing to install a third party application in their mobile devices in order to enhance their English Language proficiency. More than 87% have shown a positive response to that statement where around 10.7% have shown a neutral response. This shows that learners have a strong belief in enhancing English Language Learning through mobile. When questioning whether learners are willing to have the learning materials in their mobile devices itself, the responses were positive where more than 93% agreed to have the resources in their own devices. Further, the mobile device has been proved to be a tool which can be sued to utilize the spare time of a person. When asking about the teacher student communication through the mobile devices, 70% of the respondents have agreed the statement and 18.6% have provided a neutral response. Rest of 11.4% have provided a negative response for this statement. This shows that still there is a small portion of students who believe that the teacher student communication cannot be enhance in the mobile learning environment. Few negative statements were issued in order to find out the opponents of m-learning. 11% of the respondents believed that learning through mobile devices will not provide any advantages. Rest of 89% have provided a neutral to positive response. This shows a promising value for m-learning in the area of English Language Learning.

## 4.3 Interaction Patterns

The interaction pattern specific questions were issued only among the external undergraduates (G2), since they interact through various means. The internal undergraduates (G1) were assumed to have a similar pattern where they meet physically in a day to day manner, hence the face to face interaction among the internal students was strong.

The frequency of face to face meetings was less among the G2. It was revealed that only 4% of student meet in a daily basis. This was through the courses conducted in a physical environment. 36% of respondent stated that they meet their colleagues only few times a week. 24.5% of respondents stated that they do not meet their colleagues. Hence it was evident that the face to face interaction was in a minimal level among G2. Their main communication medium was found to be forum discussions (31.5%) and social networks (30.5%). 21.5% of students were communicating over the phone while the face to face communication was found to be in a lesser percentage (16.5%).

## 5   The need of a mobile based solution

This survey was distributed among a set of school leavers and undergraduates with the objective of identifying the prospects of m-learning in the Sri Lankan context. In first place, the ownership of mobile phone is 100% among the target group which shows a positive nature of introducing m-learning. Further, the smartphone usage among the target community is in a considerable amount, yet it is growing. Android based mobile phones seems to be more popular among the community and they are familiar with installing third party applications in their mobile phones. This shows that there is a potential to introduce an android based mobile application among the study community. Further, when analysing the data it is evident that there is more opportunity for content based learning solution with audio lessons. Further, small activity based lessons will also be accepted by the community. Further, since the community is already used to interacting through a virtual environment, a forum / discussion board will be highly accepted. As a promising value, it is evident through this survey that there is a great potential for android mobile based learning application and there is a strong potential for m-learning in the area of English Language Learning among the study community.

### 5.1   Limitations and Future research

This survey was conducted only among undergraduates and school leavers who are following a degree program (Undergraduates of UCSC and student following BIT). Hence, this study is limited to students who follow a degree program in means of online or face to face environment where the others who are not following such degree program are not considered. Further, the respondents were students who were following an IT related degree program which makes the results biased for such community. Within these limitations, we try to conduct this survey since we believe those students will be ideal for an m-learning based experiment since they already have exposure to e-learning and mobile technologies.

Through analysing the results it is evident that the study can be continued by providing an android based m-learning solution. Further, such suitable applications can be designed with the collaboration of subject matter experts in order to assure acceptance and sustainability. As this study adopts activity theory, the future research will use Activity theory again to analyse the factors affecting the context of m-Learning through .

## 6   Conclusion

Through this survey it is evident that the mobile penetration is in a satisfactory level among the study community. They are more prone towards technology and familiar with mobile applications. Majority of the learner community is between the age group of 18 – 25 which shows that they are young and having strong attitude towards m-learning. Hence, this will be a good opportunity to introduce a mobile based learning solution which will have strong acceptance among the study community.

Through this analysis the possible learning solution has been identified as an android mobile based learning application. Learner prefer a content based lessons and audio lessons. Further there is a potential for small activities as well. SMS also has been identified as a possible and accepted medium of learning but the cost constrains exists. Hence, a method of interaction also should be proposed along with the learning solution.

## 7   References

Aamri, and Suleiman, K. 2011, "The Use of Mobile Phones in Learning English Language by Sultan Qaboo University Students: Practices, Attitudes and challenges" *Canadian Journal on Scientific & Industrial Research* (2:3), December, pp. 143- 152,

Corbail & Valdes-Corbeil,2007, "Are You Ready for Mobile Learning?" *(Educase article) Educase Quartely*, November

Engeström, Y., 1987, Learning by expanding . Helsinki: Orienta-konsultit.

Gounder, 2011, "What is the potential impact of using mobile devices in higher education", *Proceedings of SIG GlobDev Fourth Annual Workshop*, December

Karunaratne, I.M., 2003, "Teaching English in urban Sri Lanka-Some pedagogical issues"

Kim, S.H., Mims, C., & Holmes, K.P. 2006,"An Introduction to Current Trends and Benefits of Mobile Wireless Technology Use in Higher Education"

Kukulska-Hulme. A, L. S, 2008, "An overview of mobile assisted language learning: From content delivery to supported collaboration and interaction", *Cambridge Univ Press*.

Kutti. K, 1995, "Activity Theory as a potential framework for human computer interaction research", Context and Consciousness: Activity Theory and Human Computer Interaction, *Cambridge: MIT Press,* pp. 17-44.

Laouris. Y & Eteokleous. N., "We need an Educationally Relevant Definition of Mobile Learning"

Lu, M., 2008, "Effectiveness of vocabulary learning via mobile phone", *Journal of Computer Assisted Learning*, (24:6), pp. 515-525.

Mwanza. D, 2001, "Where Theory meets Practice: A Case for an Activity Theory based Methodology to guide Computer System Design" *Proceedings of INTERACT' 2001: Eighth IFIP TC 13 Conference on Human-Computer Interaction*, Japan.

Pettit and Hulme, 2007, "Going with Grain: Mobile Devices in practice", *Australasian Journal of Educational Technology*, (23:1), pp.17-33.

Traxler, J., 2007, "Definign, Discussing and Evaluating Mobile Learning: The moving finger writes and having writ...", *International Review of Research and Open and Distance Learning*, (8:2)

Uden. L, 2007, "Activity theory for designing mobile learning", *International Journal on Mobile Learning and Organisation*, (1:1)

Yang, J., 2013. "Mobile Assisted Language Learning: Review of the Recent Applications of Emerging Mobile Technologies", *English Language Teaching. Canadian Center of Science and Education*.

'Design-Based Research: An Emerging Paradigm for Educational Inquiry',. 2003. *Educational Researcher* (32:1), pp. 5-8(doi: 10.3102/0013189x032001005).

Wang, F., and Hannafin, M. 2005. 'Design-based research and technology-enhanced learning environments', ETR&D (53:4), pp. 5-23(doi: 10.1007/bf02504682).

Telecommunications Regulatory Commission of Sri Lanka,. 2015. (available at http://www.trc.gov.lk; retrieved February 2, 2015).

Mobitel Sri Lanka,. 2015. http://www.mobitel.lk/ retrieved February 3, 2015.

Dialog Sri Lanka,. 2015. http://www.dialog.lk/ retrieved February 2, 2015.

British Council Sri Lanka,. 2015. http://www.britishcouncil.lk/ retrieved February 3, 2015.

Ratwatte, H. 2011. 'Why learn English? A comparative Study of Popular Belief vs. belief of educated 'masses'', in Annual Academic Sessions, Open University of Sri Lanka, pp. 195 - 199.

## Acknowledgements

The first author would like to express her gratitude to the University of Colombo School of Computing (UCSC), Sri Lanka for facilitating to conduct this research work. The authors wish to convey their sincere thanks to the National Science Foundation, Sri Lanka for the research grant. Further, thanking to all the participants who spent their time in contributing to this study.

## Copyright